\documentclass[useAMS,usenatbib]{mn2e}

\usepackage{graphicx}
\usepackage{color}
\def\ld{$\lambda_{ld}$}
\def\hd{$\lambda_{hd}$}
\def\ild{$I_{ij}({ld}$)}
\def\ihd{$I_{ij}({hd}$)}
\title[Mapping low and high-density clouds]
{Mapping low and high density clouds in astrophysical nebulae by imaging forbidden line emission\thanks{Based on observations obtained at the Gemini Observatory, which is operated by the
Association of Universities for Research in Astronomy, Inc., under a cooperative agreement
with the NSF on behalf of the Gemini partnership: the National Science Foundation (United
States), the Science and Technology Facilities Council (United Kingdom), the
National Research Council (Canada), CONICYT (Chile), the Australian Research Council
(Australia), Minist\'erio da Ci\^encia e Tecnologia (Brazil) and SECYT (Argentina).}}
\author[Steiner, Menezes, Ricci \& Oliveira]{J. E. Steiner$^{1}$\thanks{E-mail:
steiner@astro.iag.usp.br (JES)}, R. B. Menezes$^{1}$, T.V. Ricci$^{1}$
and A. S. Oliveira$^{2}$\\
$^{1}$Instituto de Astronomia, Geof\'{\i}sica e Ci\^encias Atmosf\'ericas, Universidade de S\~ao Paulo, 05508-900,
    S\~ao Paulo, SP, Brasil\\
$^{2}$IP\&D, Universidade do Vale do Para\'{\i}ba, Av. Shishima Hifumi, 2911, CEP 12244-000, S\~ao Jos\'e dos Campos, SP, Brasil}
\begin{document}

\date{Accepted . Received }

\pagerange{\pageref{firstpage}--\pageref{lastpage}} \pubyear{2008}

\maketitle

\label{firstpage}

\begin{abstract}
Emission line ratios have been essential for determining physical parameters such as gas temperature and density in astrophysical gaseous nebulae. With the advent of panoramic spectroscopic devices, images of regions with emission lines related to these physical parameters can, in principle, also be produced. We show that, with observations from modern instruments, it is possible to transform images taken from density sensitive forbidden lines into images of emission from high and low-density clouds by applying a transformation matrix. In order to achieve this, images of the pairs of density sensitive lines as well as the adjacent continuum have to be observed and combined. 

We have computed the critical densities for a series of pairs of lines in the infrared, optical, ultraviolet and X-rays bands, and calculated the pair line intensity ratios in the high and low-density limit using a 4 and 5 level atom approximation. In order to illustrate the method we applied it to GMOS-IFU data of two galactic nuclei. We conclude that this method provides new information of astrophysical interest, especially for mapping low and high-density clouds; for this reason we call it ``the ld/hd imaging method''.
\end{abstract}

\section{Introduction}

Forbidden line intensity ratios from given species (O$^{+}$, O$^{++}$, N$^{+}$, S$^{+}$) have been widely used in the literature to derive average electron temperatures and densities in astrophysical nebulae. The method for measuring the electron temperature was suggested by \citet{menzel} while the idea of using the [\mbox{O\,{\sc ii}}] line intensity ratios to measure electron densities was suggested by \citet{aller49} and worked out quantitatively by \citet{seaton54} for both [\mbox{O\,{\sc ii}}] and [\mbox{S\,{\sc ii}}] lines. An early review of these methods is given by \citet{seaton60}. Since then, intensity ratios for lines from many other species have been proposed and used.

Electron temperatures are sensitive to the intensity ratio of the auroral to the nebular components, for example, for the Carbon-like 2p$^2$ and the Silicon-like 3p$^2$ ions. A classical intensity ratio is that of [\mbox{O\,{\sc iii}}] $I\lambda4363/I\lambda5007$. Average electron densities are obtained, for example, from the Nitrogen-like 2p$^3$ and Phosphorus-like 3p$^3$ ions. Intensity ratios, used very often, are those of [\mbox{O\,{\sc ii}}] $I\lambda3726/I\lambda3729$ and [\mbox{S\,{\sc ii}}] $I\lambda6716/I\lambda6731$ lines. With the development of infrared, ultraviolet and X-ray detectors, other pairs of lines have also been used. A comprehensive review on the subject is given in \citet{osterbrock}.

In the traditional single aperture spectroscopic approach, one obtains a single spectrum of a given object and only one intensity ratio is measured. This has the obvious disadvantage of providing a single average property (temperature and/or density) for the object along the slit. With the development of two-dimensional spectroscopic devices such as the Integral Field Units (IFU) and Fabry-Perot instruments, a new approach is possible as one can obtain simultaneously the average property along the line of sight for each point of the object on the sky, if the object is spatially resolved.

In this paper we present a method of transforming two images of density-sensitive emission lines into two other images, of high and low-density cloud emission. We demonstrate the method by applying it to two objects with extended nebular emission. We will not discuss the observations in great detail nor the physics of each object. The reader is referred to the papers in which the objects are discussed, for specific information. The focus of this paper is to present the method and show some results, for illustration only.

\section{Emission from low and high-density clouds}

The pair of [\mbox{S\,{\sc ii}}] $\lambda\lambda$6716/6731 lines is, perhaps, the most used density-sensitive pair of lines. 
As all forbidden lines, these are also sensitive to the gas density. Their ratio is density-sensitive because they come from distinct upper levels.
As a consequence, the critical density for de-excitation of the $\lambda_{ld}$(6716 {\AA}) line is $N_c=1,400$ cm$^{-3}$, while for the $\lambda_{hd}$(6731 {\AA}), is $N_ c=3,600$ cm$^{-3}$. The ratio $R(\lambda_{ld}$6716/$\lambda_{hd}$6731) has a limit of $R_{ld}=1.44$ for low-density clouds ($N_e \sim 10~cm^{-3}$) and a limit of $R_{hd}= 0.44$ for high-density clouds ($N_e \sim 10^4~cm^{-3}$). In Table~\ref{lines} we present a list of lines that belong to density-sensitive configurations with respective critical densities and limit line ratios.

The two emission images of the lines, \ld, of low critical density and \hd, of high critical density, have intensities $I_{ij}(\lambda_{ld}$) and $I_{ij}(\lambda_{hd}$). These two images can be transformed in two new images \ild~ and \ihd~ of low and high-density cloud emission by the transformation

\begin{equation}
\label{matriz}
\left[ \begin{array}{c}
{I}_{ij}({ld}) \\
{I}_{ij}({hd}) \\
\end{array} \right]=\frac{1}{{R}_{ld}-{R}_{hd}} \cdot 
\left[ \begin{array}{cc}
1 & -{R}_{hd}\\
-1 & {R}_{ld} \\
\end{array} \right] \cdot
\left[ \begin{array}{c}
{I}_{ij}({\lambda_{ld}}) \\
{I}_{ij}({\lambda_{hd}}) \\
\end{array} \right]
\end{equation}

Simple algebra shows that this holds by calculating the emission properties, using the $R_{ld}$ and $R_{hd}$ from Table~\ref{lines}. It is also simple to show that the two transformed images have the property that 

\begin{equation}
\label{soma}
{I}_{ij}(\lambda_{hd}) = {I}_{ij}({ld}) + {I}_{ij}({hd})
\end{equation}

Therefore this transformation can be considered a decomposition of the image $I_{ij}(\lambda_{hd}$) into the high and low-density cloud emission images \ild~ and \ihd.

If the observed emission from a given object comes from low-density clouds only, then all the emission will be in \ild. If we have high-density clouds only, all the emission will be in image \ihd. However, in practice, we may also have clouds with intermediate densities. For these clouds part of their emission will be computed in one image and part in the other; the proportion will depend on how far the actual ratio (and density) is from the two extremes.

\begin{table*}
 \centering
 % \begin{minipage}{140mm}
  \caption{Critical and reference densities for selected lines in the infrared, optical, ultraviolet and X-rays, 
	as well as their limit line intensity ratios.   \label{lines}}
  \begin{tabular}{@{}llllllllll@{}}
\hline
  \multicolumn{10}{c}{\bf{Infrared lines:}} \\ 
\hline
Species & \ld      & \hd      & $N_c(ld)$   & $N_c(hd)$   & $R_{ld}$ & $R_{hd}$ & $N_r(ld)$  & $N_r(hd)$  & Refs.$^{\mathrm{a}}$ \\
        & ($\mu$m) & ($\mu$m) & (cm$^{-3}$) & (cm$^{-3}$) &          &          & (cm$^{-3}$)& (cm$^{-3}$)&      \\
\hline
  \multicolumn{10}{c}{C-like 2p$^2$ ions} \\ 
\mbox{N\,{\sc ii}}   & 205 & 122 & 4.4 $\times 10$   & 2.8 $\times 10^2$ & 1.44 & 0.10 & 1.8               & 2.7 $\times 10^2$& 1,2     \\
\mbox{O\,{\sc iii}}  & 88  & 52  & 4.9 $\times 10^2$ & 3.5 $\times 10^3$ & 1.78 & 0.10 & 1.2 $\times 10$   & 3.1 $\times 10^3$& 1,2     \\
\mbox{Ne\,{\sc v}}   & 24  & 14  & 6.7 $\times 10^3$ & 3.5 $\times 10^4$ & 1.10 & 0.11 & 4.3 $\times 10^2$ & 3.5 $\times 10^4$& 1,2     \\ \\
  \multicolumn{10}{c}{Si-like 3p$^2$ ions} \\ 
\mbox{S\,{\sc iii}}   & 33 & 19 & 1.4 $\times 10^3$ & 1.2 $\times 10^4$ & 2.25 & 0.08 & 2.3 $\times 10$   & 9.8 $\times 10^3$& 1,2     \\
\mbox{Ar\,{\sc v}}    & 13 & 8  & 2.9 $\times 10^4$ & 1.6 $\times 10^5$ & 1.75 & 0.13 & 8.3 $\times 10^2$ & 1.9 $\times 10^5$& 1,2     \\
\hline
  \multicolumn{10}{c}{\bf{Optical and ultraviolet lines:}} \\ 
\hline
Species & \ld      & \hd      & $N_c(ld)$   & $N_c(hd)$   & $R_{ld}$ & $R_{hd}$ & $N_r(ld)$  & $N_r(hd)$  & Refs.$^{\mathrm{a}}$ \\
        & ({\AA})  & ({\AA})  & (cm$^{-3}$) & (cm$^{-3}$) &          &          & (cm$^{-3}$)& (cm$^{-3}$)&      \\
\hline
  \multicolumn{10}{c}{N-like 2p$^3$ ions} \\ 
\mbox{N\,{\sc i}}  &5200 &5198 &6.1$\times 10^2$ &2.0$\times 10^3$ &1.49 &0.38 &3.6$\times 10$   &3.8$\times 10^3$ &1,2,3 \\
\mbox{O\,{\sc ii}} &3729 &3726 &9.4$\times 10^2$ &4.4$\times 10^3$ &1.49 &0.26 &4.5$\times 10$   &5.5$\times 10^3$ &1,2 \\
\mbox{Ne\,{\sc iv}}&2424 &2422 &1.8$\times 10^4$ &1.3$\times 10^5$ &1.50 &0.16 &7.6$\times 10^2$ &1.2$\times 10^5$ &1,2 \\
\mbox{Na\,{\sc v}} &2068 &2067 &9.5$\times 10^4$ &1.1$\times 10^6$ &1.50 &0.10 &3.8$\times 10^3$ &6.9$\times 10^5$ &1,3 \\ \\
  \multicolumn{10}{c}{P-like 3p$^3$ ions} \\ 
\mbox{S\,{\sc ii}}  &6716 &6731 &1.4$\times 10^3$ &3.6$\times 10^3$ &1.44 &0.44 &8.1$\times 10$   &5.9$\times 10^3$ &1,2 \\
\mbox{Cl\,{\sc iii}}&5517 &5537 &1.2$\times 10^4$ &6.7$\times 10^4$ &1.41 &0.21 &6.0$\times 10^2$ &6.6$\times 10^4$ &1,3 \\
\mbox{Ar\,{\sc iv}} &4711 &4740 &7.1$\times 10^4$ &4.5$\times 10^5$ &1.42 &0.18 &2.9$\times 10^3$ &3.6$\times 10^5$ &1,2 \\
\mbox{K\,{\sc v}}   &4123 &4163 &2.9$\times 10^5$ &3.8$\times 10^6$ &1.37 &0.09 &1.3$\times 10^4$ &1.8$\times 10^6$ &1,3 \\ \\
  \multicolumn{10}{c}{Be-like 2s$^2$ ions} \\ 
\mbox{C\,{\sc iii}}  &1907 &1909 &7.4$\times 10^4$ &9.7$\times 10^8$    &1.51 &7.5$\times 10^{-5}$ &2.6$\times 10^3$ &4.9$\times 10^5$ &1,2,3 \\
\mbox{N\,{\sc iv}}   &1483 &1487 &1.4$\times 10^5$ &4.4$\times 10^9$    &1.48 &3.2$\times 10^{-5}$ &4.8$\times 10^3$ &9.3$\times 10^5$ &1,2,3 \\
\mbox{O\,{\sc v}}    &1214 &1218 &3.7$\times 10^5$ &2.6$\times 10^{10}$ &1.44 &1.5$\times 10^{-5}$ &1.3$\times 10^4$ &2.5$\times 10^6$ &1,2,3 \\ \\
 \multicolumn{10}{c}{Mg-like 3s$^2$ ions} \\ 
\mbox{Si\,{\sc iii}} &1883 &1892 &4.2$\times 10^4$ &3.2$\times 10^{10}$ &1.49 &1.5$\times 10^{-6}$ &1.5$\times 10^3$ &3.1$\times 10^5$ &1,2,3 \\
\hline
  \multicolumn{10}{c}{\bf{X-ray lines:}} \\ 
\hline
Species & \ld      & \hd      & $N_c(ld)$   & $N_c(hd)$   & $R_{ld}$ & $R_{hd}$ & $N_r(ld)$  & $N_r(hd)$  & Refs.$^{\mathrm{a}}$ \\
        & ({\AA})  & ({\AA})  & (cm$^{-3}$) & (cm$^{-3}$) &          &          & (cm$^{-3}$)& (cm$^{-3}$)&      \\
\hline
  \multicolumn{10}{c}{He-like 1s$^2$ ions} \\ 
\mbox{C\,{\sc  v}}   &41,46 &40,71 &2.2$\times 10^9$    &4.4$\times 10^{15}$ &12.65&2.4$\times 10^{-6}$ &5.8$\times 10^{6}$ &9.5$\times 10^{10}$ & 1,4,5\\
\mbox{N\,{\sc vi}}   &29,53 &29,08 &1.9$\times 10^{10}$ &1.6$\times 10^{16}$ &5.20 &2.4$\times 10^{-6}$ &1.2$\times 10^{8}$ &3.3$\times 10^{11}$ & 1,4,5\\
\mbox{O\,{\sc vii} } &22,10 &21,80 &9.0$\times 10^{10}$ &4.6$\times 10^{16}$ &3.28 &2.6$\times 10^{-6}$ &1.0$\times 10^{9}$ &1.0$\times 10^{12}$ & 1,4,5,6\\
\mbox{Si\,{\sc xiii}}&6,739 &6,686 &1.1$\times 10^{14}$ &4.0$\times 10^{19}$ &1.91 &3.1$\times 10^{-6}$ &2.8$\times 10^{12}$&9.3$\times 10^{14}$ & 1,4,5,6\\

\hline
\end{tabular}
\smallskip\noindent
\begin{list}{}{}
\item[$^{\mathrm{a}}$] References: 1- \citet{nist}; 2- \citet{osterbrock}; 3- \citet{aller79}; 4- \citet{porquet}; 5- \citet{pradhan}; 6- \citet{zhang}.
\end{list}

%\end{minipage}
\end{table*}

\begin{figure}
%\vspace*{10pt}
\centerline{\includegraphics[width=90mm]{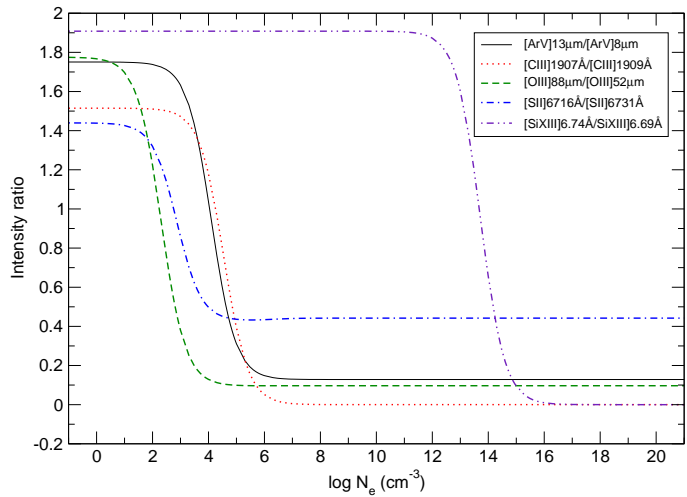}}
\caption{Calculated line ratios as functions of $N_e$ for a sample of ions. For optical and ultraviolet wavelengths, an electron temperature of $T_e=10000$ K was assumed, while for X-ray wavelength, $T_e=10^6$ K.  We have only plotted a few ions, for illustration. For practical reasons, ions with ratios larger than 2 were excluded. \label{fig1}}
\end{figure}

The question that arises here is how to define the ``intermediate densities'', as this range of densities varies from species to species (see Fig.~\ref{fig1}). We will define a ``low electron reference density'', $N_r(ld)$, as the density for which the line intensity ratio decreases the low-density ratio limit by 0.1:

\begin{equation}
{R}({I}(\lambda_{ld})/ {I}(\lambda_{hd})) = {R}_{ld} - 0.1
\end{equation}

In the same way a ``high electron reference density'', $N_r(hd)$, is defined as the density for which the line intensity ratio increases the high-density ratio by 0.1:

\begin{equation}
{R}({I}(\lambda_{ld})/ {I}(\lambda_{hd})) = {R}_{hd} + 0.1
\end{equation}

We have chosen 0.1 as the tolerance for the line ratios somewhat arbitrarily. We could have chosen, for example, 0.05. The choice has to do with observational uncertainties, not with theoretical reasons. Examining Fig.~\ref{fig1}, we see that these two choices would not produce significant differences in the reference densities. Given the typical observational errors in lines that could be frequently weak, we believe that 0.1 is a more reliable quantity than 0.05, for example.

These reference densities were also computed and are listed in Table~\ref{lines}. We will call the low-density regime when $N_e < N_r(ld)$, the intermediate-density regime when $N_r(ld) < N_e < N_r(hd)$ and the high-density regime when $N_e > N_r(hd)$. 

From the data in Table~\ref{lines} one can see that the low reference density is distinct from the low critical density.  The ratio $N_r(ld)/N_c(ld) \sim 15 - 61$ for the infrared, optical and ultraviolet lines, but presents a higher range of 39 -- 380 for the X-ray emitting lines. The high reference density is similar (within a factor 2) to the high critical density except for the Be-like and He-like ions. For these ions the ratio $N_c(hd)/N_r(hd) \sim 10^4 - 10^5$ and $ \sim 4\times10^4$ respectively.
These large numbers are related to the fact that the high density lines in these species are intercombination lines that have typically much higher critical density when compared to the corresponding forbidden line. The high reference density is, surprisingly, closely related to the low critical density for all infrared, optical and ultraviolet lines, as $N_r(hd)/N_c(ld) \sim 5.1 - 7.3$. For the X-ray lines this ratio varies from 8.4 to 43.

\section{Calculations of line intensity ratios, critical and reference densities}

Pairs of lines other than [\mbox{S\,{\sc ii}}], but from ions with the same kind of configuration, include 
[\mbox{Cl\,{\sc iii}}], [\mbox{Ar\,{\sc iv}}] and [\mbox{K\,{\sc v}}], all in the optical region of the spectrum. These are the Phosphorus-like 2p$^3$ ions. The Nitrogen-like 2p$^3$ ions include [\mbox{N\,{\sc i}}] and [\mbox{O\,{\sc ii}}] in the optical and 
[\mbox{Ne\,{\sc iv}}] and [\mbox{Na\,{\sc v}}] in the ultraviolet (Table~\ref{lines}).

In the infrared, pairs of lines of the Carbon-like 2p$^2$ and the Silicon-like 2p$^2$ are also useful density-sensitive lines. With the advancement of ultraviolet and X-ray astronomy, other types of species have been used, namely, the intensity ratio of the forbidden line to the inter-combination lines of species such as the Beryllium-like 2s$^2$ and the Helium-like 1s$^2$ ions.  Typical examples are those of the intensity of [\mbox{C\,{\sc iii}}] $I(\lambda$1907 {\AA}) to that of \mbox{C\,{\sc iii}}] $I(\lambda$1909 {\AA}) and [\mbox{O\,{\sc vii}}] $I(\lambda$22.10 {\AA}) to that of \mbox{O\,{\sc vii}}] $I(\lambda$21.80 {\AA}), (see Table~\ref{lines}).

In order to obtain the line ratios and critical densities of the ions, we obtained the populations relative to the upper energy levels involved in the transition. For this we calculated the equilibrium taking into account the collisional excitation, de-excitation and radiative decay; no recombination was considered. For the C-like, Si-like, N-like and P-like ions, we used a five level approximation. For the Be-like and Mg-like ions a four level approximation was used. For the X-ray He-like ions we used a five level approximation but without taking into account the radiative transitions within the sub-levels of the triplet. For the \mbox{C\,{\sc v}} and \mbox{N\,{\sc vi}} ions we also did not considered the collisional transitions between these sub-levels, as these constants were not to be found in the literature. References for the atomic parameters for each ion are given in Table~\ref{lines}. We only show ions for which we were able to find all atomic parameters necessary to perform all the calculations.

For the infrared, optical and ultraviolet lines we have assumed a nebular electronic temperature of $T^*=10^4$ K while for the X-ray lines we adopted $T_{\mathrm{X}}=10^6$ K. In real situations the temperature may be somewhat different; in such cases, for the infrared, optical and ultraviolet lines the following equations may be applied

\begin{equation}
N_r(T_e) = N_r(T^*)(T_e/T^*)^{\alpha} 
\end{equation}

\begin{equation}
N_c(T_e) = N_c(T^*)(T_e/T^*)^{\beta} 
\end{equation}

and

\begin{equation}
R(T_e) = R(T^*)(T_e/T^*)^{\gamma}	
\end{equation}

Values for these indices are listed in Table~\ref{indices} for all ions for which we found all relevant atomic parameters for at least two temperatures.
As can be seen in Table~\ref{indices}, the line ratios $R_{hd}(T_e)$ is temperature independent 
while $R_{ld}(T_e)$ also is nearly temperature independent, except for the infrared lines. Notice that for the X-ray emitting line ratios $R_{ld}$ and $R_{hd}$, we tabulated the ratios according to the definition of \citet{porquet}.

For the X-ray lines, however, the above equations are not a good approximation. A better description is given by

\begin{equation}
\label{fig8}
log R(T_e) = log R(T_X) + A[(T_X/T_e)^\delta -1] 
\end{equation}

and

\begin{equation}
\label{fig9}
log N(T_e) = log [N(T_X)] + B~ log T_X/T_e + C(log T_X/T_e)^2
\end{equation}

Constants for equations~\ref{fig8} and \ref{fig9} are given in Table~\ref{BCconst}. For the calculations of the constans in this table, we have not taken into account the radiative and collisional transitions between the triplet sublevels. 

Where possible, all ratios obtained were double checked using the CLOUDY program. One should note that, in our calculations, some line ratios are slightly different from those provided by the Cloudy models. Examples are the high-density limit of the [\mbox{O\,{\sc ii}}] line ratio and the low-density limit of [\mbox{S\,{\sc iii}}] and [\mbox{Ne\,{\sc v}}] infrared line ratios. For these ratios, Cloudy provides a difference of $+0.10$, $+0.25$ and $+0.15$, respectively.  Our understanding is that these few and small differences arise because of recombination contribution to these lines and, perhaps, due to different electron temperatures (see Table~\ref{indices}).

\begin{table}
 \centering
 %\begin{minipage}{140mm}
  \caption{The $\alpha$, $\beta$ and $\gamma$ indices for the ions for which data are available in the literature.\label{indices}}
  \begin{tabular}{@{}cccc@{}}
  \hline
   Ion & $\alpha (ld/hd)$ & $\beta (ld/hd)$  & $\gamma (ld/hd)$   \\
  \hline
\mbox{O\,{\sc iii}} & 0.52/0.27 & 0.34/0.30 & -0.24/-0.03 \\
\mbox{S\,{\sc iii}} & 0.82/0.32 & 0.40/0.36 & -0.24/-0.08 \\
\mbox{N\,{\sc i}}   & -0.50/-0.55 & -0.55/-0.52 & -0.01/0.00  \\
\mbox{O\,{\sc ii}}  & 0.42/0.28 & 0.35/0.44 & -0.01/0.00  \\
\mbox{Ne\,{\sc iv}} & 0.44/0.42 & 0.43/0.45 & -0.03/-0.01 \\
\mbox{S\,{\sc ii}}  & 0.54/0.29 & 0.44/0.50 & -0.03/0.00  \\
\mbox{C\,{\sc iii}} & 0.27/0.30 & 0.31/0.32 &  0.02/0.00  \\
\hline
\end{tabular}
%\end{minipage}
\end{table}

\begin{table}
 \centering
 %\begin{minipage}{140mm}
  \caption{The constants for the X-ray line property variations with temperature.\label{BCconst}}
  \begin{tabular}{@{}ccc@{}}
  \hline
Constant  & $N_c (ld/hd)$ & $N_r (ld/hd)$     \\
  \hline
$B(\mbox{C\,{\sc v}})    $ & -0.198/-0.322 & -0.386/-0.202 \\
$C(\mbox{C\,{\sc v}})    $ &  0.098/-0.020 & -0.126/0.102  \\
$B(\mbox{N\,{\sc vi}})   $ & -0.208/-0.357 & -0.435/-0.185 \\
$C(\mbox{N\,{\sc vi}})   $ &  0.102/-0.038 &-0.114/0.117   \\
$B(\mbox{O\,{\sc vii}})  $ & -0.186/-0.359 & -0.378/-0.164 \\
$C(\mbox{O\,{\sc vii}})  $ & 0.134/-0.030  & -0.079/0.164  \\
$B(\mbox{Si\,{\sc xiii}})$ & -0.147/-0.333 & -0.738/-0.053 \\
$C(\mbox{Si\,{\sc xiii}})$ & 0.107/-0.012  &-0.223/0.136   \\
\hline
Constant  & $R_{ld}/R_{hd}$ &   \\
\hline
$A(\mbox{C\,{\sc v}})       $ & 0.014/0.027 &  \\
$\delta(\mbox{C\,{\sc v}})  $ & 1.21/1.00   &  \\
$A(\mbox{N\,{\sc vi}})      $ & 0.024/0.033 &  \\
$\delta(\mbox{N\,{\sc vi}}) $ & 1.05/1.00   &  \\
$A(\mbox{O\,{\sc vii}})     $ & 0.017/0.038 &  \\
$\delta(\mbox{O\,{\sc vii}})$ & 1.15/1.00   &  \\
$A(\mbox{Si\,{\sc xiii}})    $ & 0.147/0.072 &  \\
$\delta(\mbox{Si\,{\sc xiii}})$ & 0.889/1.00 &  \\
\hline
\end{tabular}
%\end{minipage}
\end{table}

\section{Application: The nuclei of the LINER galaxy NGC 4736 and of the starburst/AGN galaxy NGC 7582}

\begin{figure*}
%\vspace*{10pt}
\centerline{\includegraphics[width=0.65\textwidth,clip]{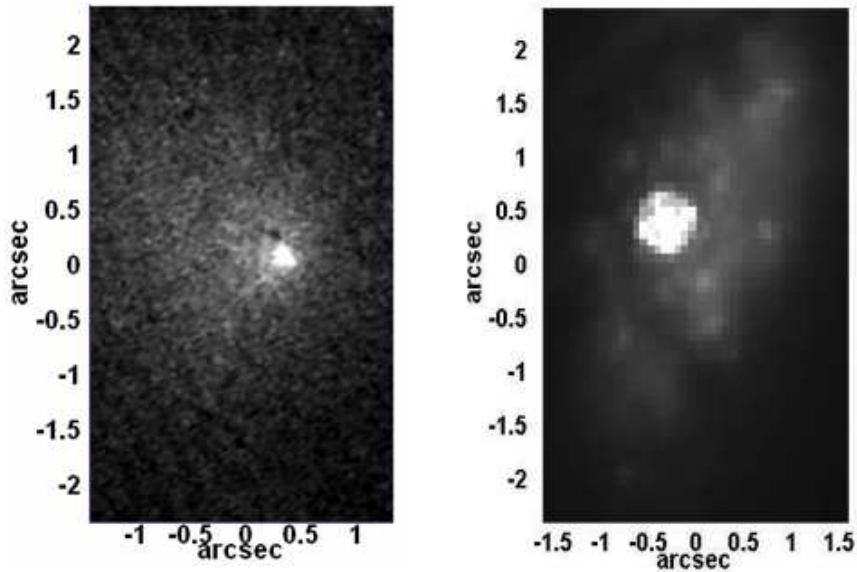}}
\caption{HST observations of the nuclear region of NGC 4736 (\textit{left,} in a UV band, F220W, centred at 2300~{\AA}) and NGC 7582 (\textit{right,} in a NIR band, F160W, centred at 16,060~{\AA}). The scale and orientation of the fields are the same as observed with the Gemini Telescope (Figs.~\ref{ngc4736} and \ref{ngc7582}). \label{hst}}
\end{figure*}

We have taken two examples of observations from the Gemini data bank to illustrate the present work. The objects analysed here are two galactic nuclei suspected for having nuclear activity. NGC 4736 is a nearby LINER galaxy while NGC 7582 is a starburst galaxy with evidence of being an AGN (Active Galactic Nucleus) from its hard X-ray emission. They were observed with the purpose of clarifying their nature, that is, to detect and characterize a hypothetical AGN. In Fig.~\ref{hst} we display HST images of the regions of interest for both galaxies.

 The data were taken with the IFU-GMOS on the Gemini telescopes. This instrument has 750 fibers in the single slit mode (500 fibers on the object and 250 on the sky) and operates in the 4000 to 11000 {\AA} spectral region. Standard procedures for  the datacube reduction were used. Before applying the transformation matrix (equation~\ref{matriz}), we have de-convolved the datacube in the spatial dimension with a Richardson-Lucy algorithm, using a gaussian PSF with FWHM equal to the observed seeing and performing 6 iterations. The goal of this paper is to show that the proposed transformation of images is a useful tool for astrophysical research; we do not intend to discuss the details of the objects themselves.

In NGC 4736, the extracted images of 
[\mbox{S\,{\sc ii}}] emission were achieved with an artificial filter matching the individual line width. An adjacent stellar continuum was subtracted. No velocity information was attempted, as the emission is weak. 

\begin{figure*}
%\vspace*{10pt}
\centerline{\includegraphics[width=0.65\textwidth,clip]{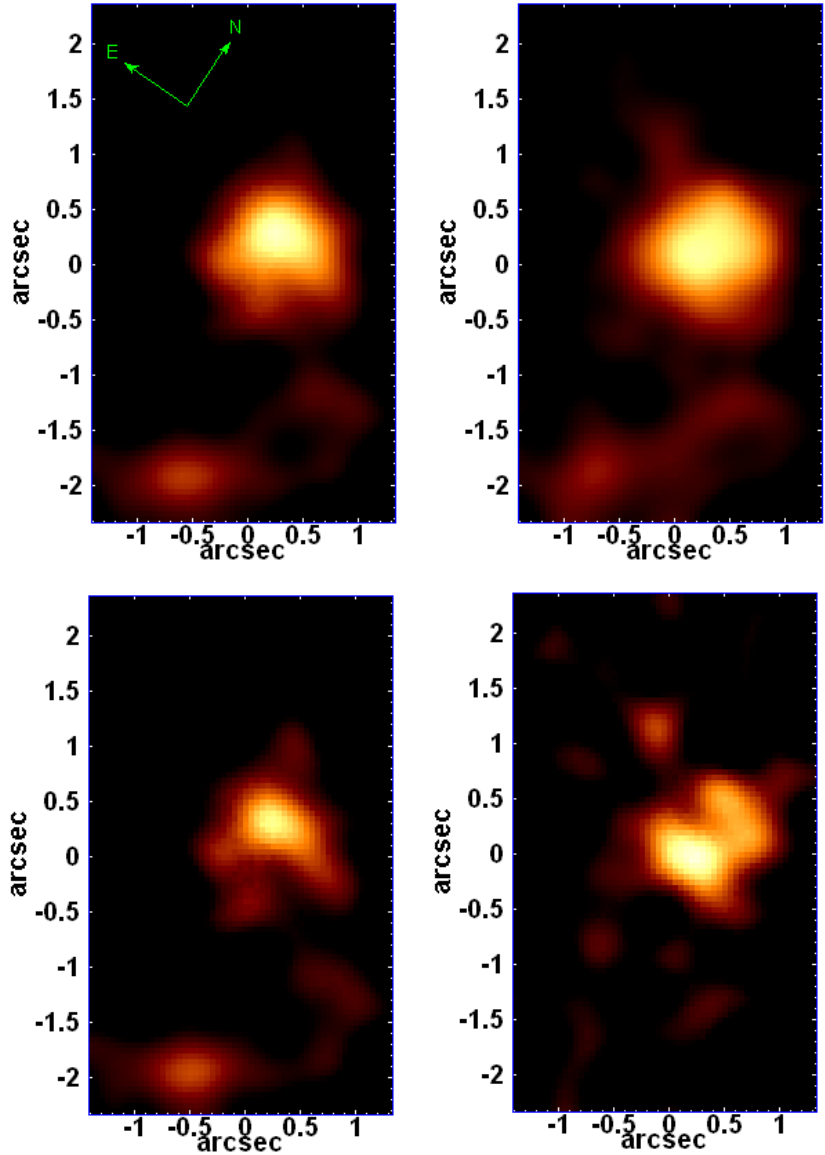}}
\caption{The nuclear region of the LINER galaxy NGC 4736: \textit{(a) Top left:} observed image of the [\mbox{S\,{\sc ii}}] emission  $I_{ij}(\lambda_{ld}$ 6716~{\AA}); \textit{(b) Top right:} observed image of the [\mbox{S\,{\sc ii}}] emission $I_{ij}(\lambda_{hd}$ 6731~{\AA}); \textit{(c) Bottom left:} transformed image of low-density cloud emission; \textit{(d) Bottom right:} transformed image of high-density cloud emission. \label{ngc4736}}
\end{figure*}

The images show condensations of low (Fig.~\ref{ngc4736}c) and high (Fig.~\ref{ngc4736}d) density gas emission. These two maps form the NLR (Narrow Line Region) of the AGN. In this case, the location of the BLR (Broad Line Region) is known to be at $x=+0.2$ and $y=+0.2$ \citep{steiner}. Two blobs of high-density clouds, separated by 0.6 arcsec are seen near the BLR, while only one is seen in the low-density cloud map. A second, but much weaker blob is seen 2.4 arcsec to the South.

Interestingly, the double structure that seems to emerge in the $hd$ image resembles the double radio source observed by \citet{kording}. The radio sources are separated by about 0.9 arcsec and have an alignment with a position angle of about $-43^{\circ}$. This can not be associated to the two blobs in hd image as they present a position angle of PA $= -10^{\circ}$.

\begin{figure*}
%\vspace*{10pt}
\centerline{\includegraphics[width=0.65\textwidth,clip]{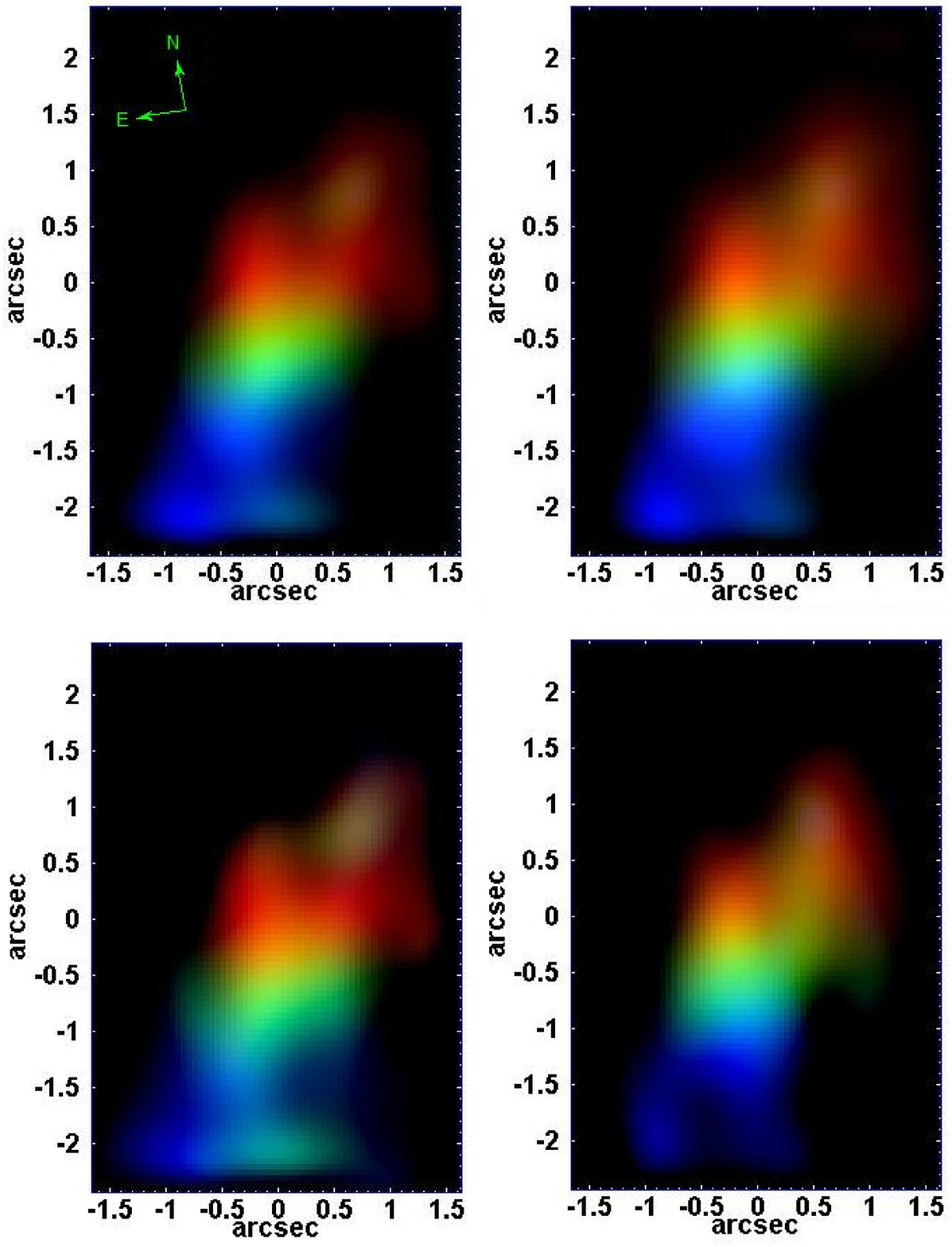}}
\caption{The nuclear region of the AGN/starburst galaxy NGC 7582: \textit{(a) Top left:} observed image of the [\mbox{S\,{\sc ii}}] emission  $I_{ij}(\lambda_{ld}$ 6716~{\AA}); \textit{(b) Top right:} observed image of the [\mbox{S\,{\sc ii}}] emission $I_{ij}(\lambda_{hd}$ 6731~{\AA}); \textit{(c) Bottom left:} transformed image of low-density cloud emission; \textit{(d)  Bottom right:} transformed image of high-density cloud emission.
The RGB colors have the following ranges: R: from 0 to 80 km s$^{-1}$; G: from 0 to $-80$ km s$^{-1}$ and B: from $-80$ to $-243$ km s$^{-1}$. Image \textit{(d)} is dominated by two \mbox{H\,{\sc ii}} regions centred at $(x=0.7; y=0.5)$ and ($-0.2; -0.5$) while image \textit{(c)} is dominated by 
the H II regions, the ionization cone and a few hot spots of unidentified origin, perhaps SNRs.
\label{ngc7582}}
\end{figure*}

In the datacube from the nucleus of NGC 7582 (Fig.~\ref{ngc7582}), the [\mbox{S\,{\sc ii}}] line emission is significantly stronger. Therefore we extracted the images with a similar procedure as above, but with narrower filters (co-adding images of contiguous wavelength pixels), so that velocity information is also registered. Three velocity bands were defined. Red is associated with a filter from 0 to $+80$ km s$^{-1}$; green is a filter from 0 to $-80$ km s$^{-1}$ and blue is from $-80$ to $-243$ km s$^{-1}$. The blue-shifted band is necessary because of the blue wing seen in most of the forbidden lines.

The images display, first of all, a clear (known) galactic rotation map \citep{morris}. In addition, in the high-density image (Fig.~\ref{ngc7582}d) the bulk of emission comes from two \mbox{H\,{\sc ii}} regions. These two \mbox{H\,{\sc ii}} regions are associated to the starburst character of this galaxy and their emission lines overshine that of the AGN. 

The low-density emission image has a complex and spatially extended structure. This complexity is due to knots that correspond to H II regions and a few hot spots, presumably Supernovae Remnants -- SNR; in addition, a diffuse bluish emission that comes from a wind, presumably photoionized by the AGN but not as much obscured, delineating the ionization cone \citep{morris, storchi}.
A paper with a full discussion of the datacube of this object is in preparation.

\section{Conclusions}

The density sensitive emission maps should not be confused with maps of line ratios $I(\lambda6716/\lambda6731)$, which are also used in the literature. Such an image does not indicate the intensity and is frequently very noisy when the denominator becomes small or even zero.

The reader should be cautioned to the fact that if a given spatial pixel coincides with a low-density cloud and a high-density cloud, projected along the same line of sight, one may be confused as this could be interpreted as emission coming from an intermediate-density cloud. If velocity separation is possible, as we show in the example above, then the chance for the two clouds to coincide in spatial projection and simultaneously in velocity, diminishes. 

Although the sum of the images of high-density and low-density is mathematically equal to the image of the line of highest critical density (see equation~\ref{soma}), this is not always obvious in the displayed images because of distinct LUT (look-up table) dynamical ranges and choices.

For the present methodology to be applicable, the cloud velocity distribution must be smaller than half the separation of the two lines, otherwise there will be superposition of emission on the same spectral pixel. This is usually the case for planetary nebular, 
\mbox{H\,{\sc ii}} regions, symbiotic stars etc., but may not be the case for novae and supernova remnants and most of the active galactic nuclei. It is worth mentioning that the various ionic configurations have very distinct line separations. For instance, the infrared lines are quite separated (in velocity) as are the X-ray lines (whose separation is $\sim 5000$ km s$^{-1}$) where this warning does not apply. But for the nitrogen-like 2p$^3$ configurations, the line separations are typically only 2 {\AA}. This is the case for [\mbox{O\,{\sc ii}}] $\lambda3736/\lambda3729$. In such situations the technique applies only if the involved velocities are smaller than $\sim 100$ km s$^{-1}$.

Two-dimensional optical devices such as Integral Field Units (IFUs) and Fabry-Perot have advanced significantly in recent years; the method proposed in this paper, when combined with such instruments, may provide a powerful tool for science. Future developments may extend this method to other wavebands, so that many of the pairs of lined listed in Table~\ref{lines} could become potentially useful.

We conclude that this ``ld/hd imaging method'' may be useful in mapping low and high-density cloud distribution in astrophysical nebulae. In addition to density maps it also provides a density-related intensity image in specific forbidden lines.

\section*{Acknowledgments}
We would like to thank FAPESP -- Funda\c c\~ao de Amparo \`a Pesquisa do Estado de S\~ao Paulo -- for financial support under grants 06/05203-3 and 05/03323. We would also like to thank G. Ferland for the use of the CLOUDY program and K. Taylor and R. Lopes de Oliveira for their careful reading of the manuscript.

\label{lastpage}

\end{document}